\documentclass[12pt,epsfig]{article}  
\usepackage{amsmath}
\usepackage{amssymb}
\usepackage{graphicx}
\usepackage{commath}
\usepackage[left=2.0cm,right=2.0cm, top=2.5cm,bottom=2.5cm]{geometry}
\usepackage{cite}

\begin{document}

\newcommand{\sheptitle}
{An Algebraic Analysis of Neutrino Masses and Mixings
 and Its Implications on $\mu-\tau$ Symmetric Mass Matrix}
\newcommand{\shepauthor}
{Chandan Duarah \footnote{ E-mail: chandan.duarah@gmail.com}}
\newcommand{\shepaddress}
   { Department of Physics, Dibrugarh University,
               Dibrugarh - 786004, India }
\newcommand{\shepabstract}
{We diagonalize Majorana neutrino mass matrix with the help of PMNS matrix and obtain
analytical relations between the mass matrix elements and mixing parameters,
viz., three mixing angles- $\theta_{12}$, $\theta_{23}$, $\theta_{13}$ and Dirac
CP phase $\delta$. We analyse our results in a special $\mu-\tau$ symmetric 
mass matrix which corresponds to maximal atmospheric mixing ($\theta_{23}=\pi/4$) 
and maximal CP violation ($\delta=-\pi/2$). The analysis shows that a deviation
of $\theta_{23}$ from its maximal value can be correlated with the prediction
of other two mixing angles.    \\
 
Key-words: Neutrino mass matrix, mixing matrix, $\mu-\tau$ symmetry.\\
\indent PACS number: 14.60 Pq}
\begin{titlepage}
\begin{flushright}
\end{flushright}
\begin{center}
{\large{\bf\sheptitle}}
\bigskip\\
\shepauthor
\\
\mbox{}\\
{\it\shepaddress}\\
\vspace{.5in}
{\bf Abstract}
\bigskip
\end{center}
\setcounter{page}{0}
\shepabstract
\end{titlepage}


\section{Introduction}

Precise measurements of neutrino mixing angles and mass squared differences have
been made almost possible by the continuous effort provided by the oscillation 
experiments \cite{dbay-13,reno-13,dchoo-13,icube-23,nova-atm}. Other issues like the 
determination of CP phase and mass ordering, problem of octant degeneracy are 
expected to be solved in the near future \cite{t2k-cp,nova-morder}. People are 
trying to understand the underlying theory behind the lepton flavor mixing \cite{flavor,flavor1}
using different approaches including $\mu-\tau$ symmetry \cite{mt,mt1,mt2,dsym}. 
The lepton mixing matrix or PMNS matrix, which appears in the weak charged current
of left-handed charged lepton fields and neutrino fields, can be expressed as
$U_{PMNS}=(U_L^l)^{\dagger} U_L^{\nu}$. Here $U_L^l$ and $U_L^{\nu}$ are the
mixing matrices for left handed charged leptons and neutrinos respectively. These mixing matrices also 
diagonalize the mass matrices for respective fields. In the basis where charged lepton 
mass matrix is taken as a diagonal matrix (that means $U_L^l$ is identity matrix),
lepton mixing matrix $U_{PMNS}$ and the neutrino mixing matrix $U_L^{\nu}$ become
same. The effective Majorana neutrino mass term is given by 
\begin{equation}
\mathcal{L}_{mass}=- \frac{1}{2} \overline{\begin{pmatrix}
                                  \nu_e^c & \nu_{\mu}^c & \nu_{\tau}^c \\
                                  \end{pmatrix}_L} M_{\nu} 
                                  \begin{pmatrix}
                                  \nu_e \\
                                  \nu_{\mu} \\
                                  \nu_{\tau} \\
                                  \end{pmatrix}_L,
\end{equation}
where $\nu_{\alpha}^c = C \bar{\nu_{\alpha}}^T$ represents charge conjugation of the
fields $\nu_{\alpha}$ ($\alpha=e, \mu, \tau$). $M_{\nu}$ is the Majorana neutrino mass
matrix which is a symmetric matrix having six complex elements or twelve real 
parameters. One can diagonalize this mass matrix with the help of the neutrino mixing 
matrix $U_L^{\nu}$ to obtain the three mass eigenvalues. We consider the basis where
the charged lepton mass matrix is already diagonal, such that $U_L^{\nu}$ and $U_{PMNS}$
are same. \\

In the standard parametrization PMNS matrix is parametrized as
\begin{equation}
U_{PMNS}=R_{23} U_{13} R_{12} P,
\end{equation}
 where 
\begin{equation*}
R_{12}= \begin{pmatrix}
            c_{12} & s_{12} & 0 \\
           -s_{12} & c_{12} & 0 \\
              0    &    0   & 1 \\

                \end{pmatrix}, \ \ R_{23}= \begin{pmatrix}
                                         1      & 0      & 0 \\
                                        0       & c_{23} & s_{23} \\
                                           0    & -s_{23} & c_{23} \\

                                             \end{pmatrix},
\end{equation*}
\begin{equation*}
U_{13}= \begin{pmatrix}
            c_{13} & 0     & s_{13} e^{-i \delta} \\
               0    & 1     & 0 \\
     -s_{13} e^{i \delta}  &    0   & c_{13} \\

                \end{pmatrix},
\end{equation*}
and $P=diag( e^{-i \alpha}, e^{-i \beta},1)$. Here $c_{ij}=\cos \theta_{ij}$ and
$s_{ij}=\sin \theta_{ij}$ ($i,j=1,2,3$), $\theta_{ij}$ being the mixing angles. 
$U_{13}$ contains the Dirac phase $\delta$ and $P$ contains the Majorana phases 
$\alpha$ and $\beta$. One can diagonalize $M_{\nu}$ with the help of $U_{PMNS}$
and may obtain useful relations between the mass matrix elements and mass eigenvalues
and the mixing parameters viz. three mixing angles and Dirac CP phase. The analysis
would involve twelve real quantities that parameterize $M_{\nu}$. As a simpler choice, 
instead of $M_{\nu}$, a hermitian matrix
$M=M_{\nu}^{\dagger} M_{\nu}$ can be taken for analysis. Under such simplification
number of real parameters of the mass matrix reduces to nine. Detail analysis with 
such a hermitian mass matrix is found in References \cite{masaki,masaki1,masaki2,xing,jora}. 
So far the hermitian matrix $M$ is considered, the diagonalization usually provides 
relationships between the mixing parameters and elements of $M$. Direct connections 
between mixing parameters and the elements of $M_{\nu}$ become no longer visible. 
In this work we attempt to analyze the relationships
between mixing parameters and the elements of $M_{\nu}$  by diagonalizing
 $M_{\nu}$ directly with $U_{PMNS}$. In section 2 we derive some useful relations
and obtain expressions for mixing parameters. In section 3 we analyze our results
for $\mu-\tau$ symmetric mass matrix and discuss its consequences. Section 4 is
devoted to summary and discussion. \\


\section{Mass Matrix Elements and Mixing Parameters}
We write the symmetric Majorana neutrino mass matrix as 
\begin{equation}
M_{\nu} = \begin{pmatrix}
                   A & B & C \\
                    B & D & E \\
                    C & E & F \\
       \end{pmatrix},
\end{equation}
where all the elements ($A$ to $F$) are complex. It can be diagonalized
by the PMNS matrix as 
\begin{equation}
U_{PMNS}^T M_{\nu} U_{PMNS} = M_d,
\end{equation}
where $M_d= Diag\{ M_1, M_2, M_3\}$. Since $M_{\nu}$ is a symmetric matrix the 
diagonalization (Eq.(4)) will in general lead to three complex mass eigenvalues- $M_1$, $M_2$ 
and $M_3$, which contain three unphysical phases. In Ref. \cite{xing},
the authors have discussed a parameter mismatching problem which is often encountered
in the diagonalization of $M_{\nu}$. From this point of view we note that with these 
three complex eigenvalues, parameter counting of either side of Eq.(4) is balanced. 
That means $M_{\nu}$ and $U_{PMNS}^{*}M_d U_{PMNS}^{\dagger}$ contain twelve real parameters
each. Square of the mass eigenvalues will be given by 
$m_i^2 = M^{*}_i M_i$ ($i=1,2,3$), which are free from the unphysical phases. \\

To carry out the diagonalization, we follow the useful steps available 
in Refs. \cite{xing, jora}, and re-express Eq.(4) as
\begin{equation}
 R_{23}^T M_{\nu} R_{23} = U^{*}_{13}R_{12}P^{*} M_d P^{\dagger}R_{12}^T U^{\dagger}_{13}.
\end{equation}
Denoting left hand side of Eq.(5) by $X$, matrix elements are given by
\begin{align}
X_{11} & =  A  \nonumber \\
X_{12} & =  c_{23}B - s_{23}C   \nonumber\\
X_{13} & = s_{23}B + c_{23}C    \nonumber\\
X_{22} & = c_{23}^2D + s_{23}^2 F - 2s_{23} c_{23}E    \nonumber\\
X_{23} & = s_{23}c_{23} (D - F) + (c_{23}^2 - s_{23}^2)E    \nonumber \\
X_{33} & = s_{23}^2D + c_{23}^2 F + 2s_{23}c_{23}E
\end{align}
Similarly we can obtain matrix elements of the right hand side of Eq.(5) which 
contain the mass eigenvalues $M_i$'s. Then equating the matrix  
elements of Eq.(5) we get total six equations. Half of these six equations 
provide us to express $M_i$'s in terms of $A$, $B$ and $C$,
while the other half can be solved to express $M_i$'s in terms of $D$, $E$ and $F$. 
The obtained expressions for $M_i$'s in terms of $A$, $B$ and $C$, are
\begin{align}
M_1 e^{-2i\alpha} & = A - \frac{t_{12}}{c_{13}}X_{12}
                           - t_{13} X_{13}e^{i\delta},  \nonumber \\
M_2 e^{-2i\beta} & = A + \frac{1}{c_{13}t_{12}} X_{12}
                      - t_{13} X_{13} e^{i\delta},      \nonumber \\
           M_3  & = \left( A e^{-i\delta} + \frac{1}{t_{13}} X_{13}\right) e^{-i\delta},  
\end{align}
and those in terms of $D$, $E$ and $F$ are given by
\begin{align}
M_1 e^{-2i\alpha} & = X_{22} + \frac{1}{s_{13}t_{12}} X_{23} e^{i\delta},  \nonumber \\
M_2 e^{-2i\beta}  & = X_{22} - \frac{t_{12}}{s_{13}} X_{23} e^{i\delta}, \nonumber \\
           M_3   & =  \frac{1}{c_{13}^2} X_{33} - t_{13}^2 X_{22} e^{-2i\delta}
                      - \frac{2s_{13}}{c_{13}^2} \tan 2\theta_{12} X_{23} e^{-i\delta},
\end{align}
where $t_{ij}$ ($i,j=1,2,3$), represents $\tan\theta_{ij}$ and $X_{ij}$'s ($i,j=1,2,3$),  
are defined in Eq.(6). Since the eigenvalues $M_i$'s involve unphysical phases 
we first want to eliminate them to carry out the analysis. Comparing Eq.(6)
and Eq.(7) we can easily execute the elimination.
Appearance of the Majorana phases $\alpha$ and $\beta$ in Eqs.(6) and (7) is also 
interesting to note. While eliminating $M_i$'s, these phases also disappear from
the resulting expressions which, in deed, reflects the fact that in the analysis
of the structure of $M_{\nu}$, Majorana phases are irrelevant \cite{masaki}.
Eliminating $M_i$'s we obtain three constraint equations which involve only the 
mass matrix elements and mixing parameters. These are given by
\begin{align}
D+F & = (1+e^{-2i\delta})A 
             + \left( \frac{1}{2}\sin 2\theta_{13}e^{-i\delta} - 2t_{13}\cos\delta \right) X_{13} \nonumber \\
              & \indent \quad \quad + 2\tan 2\theta_{12}
                   \left( \frac{1}{c_{13}} + c_{13}e^{-2i\delta} \right) X_{12}, \\
D-F & = -2 \left( \tan 2\theta_{23} E 
                  + \frac{t_{13}}{\sin 2\theta_{23}}X_{12}e^{-i\delta} \right) , \\
E & = \frac{1}{2}\sin 2\theta_{23} (-1+e^{-2i\delta})A 
         + \frac{1}{2}\sin 2\theta_{23} 
             \left( t_{13}e^{i\delta} +\frac{2}{\tan 2\theta_{13}}e^{-i\delta} \right) X_{13} \nonumber \\
     &  \indent \quad \quad - \frac{\sin 2\theta_{23}}{c_{13}}
                    \left( \frac{1}{\tan 2\theta_{12}} + \frac{s_{13}}{\tan 2\theta_{23}}e^{-i\delta} 
                          + \frac{2s_{13}^2}{\tan 4\theta_{12}} e^{-2i\delta} \right) X_{12}.
\end{align}
Though it will be a lengthy calculations, still one can use these equations
to express three of the mass matrix elements in terms of the others. These equations 
will be useful in the texture analysis of the mass matrix. We then proceed to express 
the mixing angles and the CP phase in terms of the mass matrix elements. To make 
the expressions simple we first want to define five quantities:
$\lambda_1 = Re(A^{*}X_{22})$, $\lambda_2 = Re(A^{*}X_{12})$,
$\lambda_3 = Re(A^{*}X_{13}e^{i\delta})$, $\lambda_4 = Re(X_{22}^{*}X_{13}e^{i\delta})$,
 $\lambda_5 = Re(X_{12}^{*}X_{13}e^{i\delta})$. Mixing angles $\theta_{13}$ 
 and $\theta_{12}$ can be expressed as \\
\begin{equation}
\tan\theta_{13} =  \abs{ - \frac{X_{23}}
                                    {X_{12}} e^{i\delta} } 
                =  \frac{\abs{X_{23}}}{\abs{X_{12}}},   
\end{equation}
\begin{eqnarray}
\tan 2\theta_{12} & =& \abs{ -\frac{2}{c_{13}} \frac{X_{12}}
                      {A -X_{22}- t_{13}X_{13}e^{i\delta}} }    \nonumber \\
                & =& \frac{2}{c_{13}}\frac{\abs{X_{12}}}
                           {\sqrt{ \abs{A}^2 + \abs{X_{22}}^2 + t_{13}^2\abs{X_{13}}^2
                            -2c_{13}\lambda_1 -2s_{13}\lambda_3 +2 t_{13}\lambda_4 }},
\end{eqnarray}
with
\begin{align}
\abs{X_{12}}^2 & = c_{23}^2 \abs{B}^2 + s_{23}^2 \abs{C}^2- \sin 2\theta_{23}Re(B^{*}C), \nonumber \\
\abs{X_{13}}^2 & = s_{23}^2 \abs{B}^2 + c_{23}^2 \abs{C}^2 + \sin 2\theta_{23}Re(B^{*}C), \nonumber \\
\abs{X_{23}}^2 & = \frac{1}{4}\sin^2 2\theta_{23} \abs{(D-F)}^2 
                      + \cos^2 2\theta_{23} \abs{E}^2 
                                + \frac{1}{2}\sin 4\theta_{23}Re((D-F)^{*}E),  \nonumber \\
\abs{X_{22}}^2 & = c_{23}^4 \abs{D}^2 + s_{23}^4 \abs{F}^2 + \sin ^22\theta_{23}\abs{E}^2   \nonumber \\
                & \indent + \frac{1}{2} \sin 2\theta_{23} \left[\sin 2\theta_{23}Re(D^{*}F)
                               -4 \left( s_{23}^2 Re(F^{*}E) +  c_{23}^2 Re(D^{*}E)\right)  \right].
\end{align}
We note that the expression for $\theta_{13}$ (Eq.(12)) is free from the mixing angle
$\theta_{12}$. The CP phase can be expressed as 
\begin{equation}
\tan\delta = \frac{Im(X_{23}^{*}X_{12})}{Re(X_{23}^{*}X_{12})}.
\end{equation}
Finally square of the mass eigenvalues can be obtained either from Eq.(7) or Eq.(8).
From Eq.(7) we have
\begin{align}
m_1^2 & = \abs{A}^2 + \frac{t_{12}^2}{c_{13}^2}\abs{X_{12}}^2 + t_{13}^2\abs{X_{13}}^2 
            - \frac{2t_{12}}{c_{13}}\left( \lambda_2 + 
                          \frac{s_{13}}{t_{12}}\lambda_3 - t_{13}\lambda_5 \right),   \nonumber \\
m_2^2  & = \abs{A}^2 + \frac{1}{t_{12}^2c_{13}^2}\abs{X_{12}}^2 + t_{13}^2\abs{X_{13}}^2 
            + 2 \frac{1}{t_{12}c_{13}}\left( \lambda_2 
                             - s_{13}t_{12}\lambda_3 - t_{13}\lambda_5 \right),   \nonumber \\
m_3^2  & = \abs{A}^2 + \frac{1}{t_{13}^2}\abs{X_{13}}^2 + \frac{2}{t_{13}}\lambda_3 ,
\end{align}
which involve matrix elements $A$, $B$ and $C$ only. Similar expressions can also be obtained
from Eq.(8) which will involve matrix elements $D$, $E$ and $F$ only. 

\section{Mass matrix with $\mu-\tau$ symmetry}
We now want to analyze our results for a special texture of mass matrix.
The general $\mu-\tau$ symmetric mass matrix has the form
\begin{equation}
\begin{pmatrix}
A & B & -B \\
B & D & E \\
-B & E & D \\
\end{pmatrix}.
\end{equation}
Here $A$, $B$, $D$ and $E$ are in general complex. The special property of $\mu-\tau$ 
symmetric mass matrix is that it predicts maximal atmospheric mixing ($\theta_{23}=\pi/4$), 
leaving $\theta_{12}$ arbitrary. In addition, above mass matrix sets $\theta_{13}=0$ and
CP phase $\delta=0$. Specific examples of such $\mu-\tau$ symmetric
mixing schemes are Bi-maximal mixing \cite{bm}, Tri-bimaximal mixing \cite{tbm} etc. 
An extended version of $\mu-\tau$ symmetric mass matrix which accommodate non-zero 
$\theta_{13}$ along with maximal CP violation is given by
\begin{equation}
\begin{pmatrix}
A & B & -B^{*} \\
B & D & E \\
-B^{*} & E & D^{*} \\
\end{pmatrix} = \begin{pmatrix}
                      A & Re(B)+iIm(B) & -\left( Re(B)-iIm(B)\right)  \\
                      \cdot & Re(D)+iIm(D) & E \\
                      \cdot & \cdot & Re(D)-iIm(D) \\
                       \end{pmatrix}.
\end{equation}
Here $A$ and $E$ are real. This mass matrix corresponds to $\delta=-(\pi/2)$. It has been 
discussed in a number of works with discrete flavour symmetry models 
\cite{maxml,maxml1,maxml2,maxml3,maxml4,maxml5}. As the oscillation data points out
a near maximal value of atmospheric mixing, it enriches the phenomenological
importance of the mass matrix. \\

For this mass matrix (with $\theta_{23}=\pi/4$ and $\delta=-(\pi/2)$), we then obtain
the other two mixing angles from Eq.(12) and Eq.(13) which are given by 
\begin{equation}
\tan \theta_{13} = -\frac{1}{\sqrt{2}}\frac{Im(D)}{Re(B)},
\end{equation} 
and
\begin{equation}
\tan 2\theta_{12} = -\frac{2}{c_{13}}\frac{\sqrt{2}\left( Re(B)\right)^2 }
                         {(A+E)Re(B)+ Im(B)Im(D)-Re(B)Re(D)}.
\end{equation}  
These expressions show that the mixing angles are purely expressible in terms of the mass
matrix elements, for the $\mu-\tau$ symmetric mass matrix in Eq.(18). \\

We now want to break the symmetry of the mass matrix in order to have deviation
of $\theta_{23}$ from its maximal value, keeping $\delta$ fixed at $-(\pi/2)$.
In order to do that we choose the positive variable $\delta\theta_{23}$ 
as a symmetry breaking parameter, where $\delta\theta_{23}$ accounts for the 
deviation of $\theta_{23}$ from its maximal value through the relation 
$\theta_{23}=(\pi/4) \pm \delta\theta_{23}$. The '$\pm$' signs correspond to the case
 whether $\theta_{23}$ lies in first or second octant respectively. A specific choice
for breaking $\mu-\tau$ symmetry then follows the mass matrix
\begin{equation}
\begin{pmatrix}
A & Re(B)\Delta_{\mp}+iIm(B)\Delta_{\pm} & -\left(Re(B)\Delta_{\pm}-iIm(B)\Delta_{\mp}\right)  \\
\cdot & Re(D) \pm E s2_{\delta\theta} + iIm(D) c2_{\delta\theta} 
               & E c2_{\delta\theta} \mp i Im(D) s2_{\delta\theta} \\
\cdot & \cdot & Re(D) \mp E s2_{\delta\theta} - iIm(D) c2_{\delta\theta}  \\
\end{pmatrix},
\end{equation}
where we have used the abbreviations- 
$\Delta_{\pm}=\cos\delta\theta_{23} \pm \sin\delta\theta_{23} $ and 
$s2_{\delta\theta}= \sin(2\delta\theta_{23})$, $c2_{\delta\theta}= \cos(2\delta\theta_{23})$.
Comparison of Eq.(18) and Eq.(21) reflects the pattern of breaking the symmetry.
The corresponding mixing matrix will look like (from Eq.(2))
\begin{equation}
\begin{pmatrix}
c_{12}c_{13} & s_{12}c_{13} & i s_{13}  \\
\frac{1}{\sqrt{2}}\left(-s_{12}\Delta_{\mp} + i c_{12}s_{13}\Delta_{\pm} \right)  
                    & \frac{1}{\sqrt{2}}\left(c_{12}\Delta_{\mp} + i s_{12}s_{13}\Delta_{\pm} \right) 
                                   & \frac{1}{\sqrt{2}}c_{13}\Delta_{\pm} \\
\frac{1}{\sqrt{2}}\left(s_{12}\Delta_{\pm} + i c_{12}s_{13}\Delta_{\mp} \right)  
                    & \frac{1}{\sqrt{2}}\left(-c_{12}\Delta_{\pm} + i s_{12}s_{13}\Delta_{\mp} \right) 
                                   & \frac{1}{\sqrt{2}}c_{13}\Delta_{\mp} \\
\end{pmatrix}.
\end{equation}
Under this symmetry breaking pattern the mixing angles $\theta_{13}$ and $\theta_{12}$ 
(from Eqs.(12) and (13)) become
\begin{equation}
\tan\theta_{13} = \frac{1}{\sqrt{2}} 
                 \sqrt{ \frac{E^2 s2_{\delta\theta}^2 + \left(Im(D)\right)^2 c2_{\delta\theta}^2}
                   {\left(Re(B)\right)^2 c_{\delta\theta}^2 + \left(Im(B)\right)^2 s_{\delta\theta}^2}},
\end{equation}
\begin{equation}
\tan 2\theta_{12} = \frac{\sqrt{2}}{c_{13}} 
              \sqrt{  \frac{\left(Re(B)\right)^2 c_{\delta\theta}^2
                           + \left(Im(B)\right)^2 s_{\delta\theta}^2}
                                 {W_1^2 + W_2^2}  },
\end{equation}
with
\begin{equation*}
 W_1=  A -Re(D) -\sqrt{2}t_{13}Im(B)c_{\delta\theta}
                                             + E c2_{\delta\theta}, \
W_2= \pm 2\left(\sqrt{2}t_{13}Re(B)s_{\delta\theta} 
                              + Im(D)s2_{\delta\theta} \right).
\end{equation*}
The abbreviations $s_{\delta\theta}/c_{\delta\theta}$ respectively represent 
$\sin\delta\theta_{23}/\cos\delta\theta_{23}$. Comparison of Eqs.(19) and (20) 
with Eqs.(23) and (24) shows the effect of $\delta\theta_{23}$ on the prediction
of mixing angles upon symmetry breaking. From the expressions (23) and (24) we
note that the mixing angles are correlated with the deviation parameter. In conclusion
these results can be viewed as follow: the mass matrix in Eq.(18) predicts maximal 
$\theta_{23}$ mixing and maximal CP violation. When we employ the parameter 
$\delta\theta_{23}$ to deviate $\theta_{23}$ from its maximal value, other two mixing
angles ($\theta_{13}$ and $\theta_{12}$) become solely dependent on $\delta\theta_{23}$, 
provided the mass matrix elements are being treated as free parameters. This result 
may have implications on model building purpose.

\section{Summary and discussion}
Majorana neutrino mass matrix, which contains six independent complex elements, can be 
diagonalized by the PMNS matrix. The diagonalization allows us to obtain relations 
between the mass matrix elements and mixing parameters. We derive three such equations 
which can be used to constrain three of the six independent mass matrix elements. 
We also find out expressions for the mixing angles and the Dirac CP phase in terms
of those mass matrix elements. These expressions and constrain equations may be
helpful in discussing specific neutrino mass models. We implicate our results on
a specific $\mu-\tau$ symmetric mass matrix that predicts $\theta_{23}=\pi/4$ and 
$\delta=-(\pi/2)$. For the mass matrix other two mixing angles are found to be 
related with the mass matrix elements by simple formulae. Next we consider a specific
pattern of broken $\mu-\tau$ symmetry of the mass matrix by a parameter that itself 
accounts for a deviation of $\theta_{23}$ from its maximal value. The case of maximal 
CP violation is remained unchanged. Under this symmetry breaking scheme the expressions 
for $\theta_{13}$ and $\theta_{12}$ show a correlation with the deviation parameter.
This, in other sense, reveal a property of neutrino masses and mixings that if we 
rely on $\mu-\tau$ symmetry to realize neutrino mixing, the mixing angles $\theta_{13}$ 
and $\theta_{12}$ can be made predictable from the deviation parameter itself, provided
the mass matrix elements are being considered as free parameters. The analysis made
in this work will help in the discussion of various models of neutrino masses 
and mixings with CP violation.


\begin{thebibliography}{23}

\bibitem{dbay-13} F. P. An et al. (Daya Bay Collab.), Phys. Rev. D 95, 072006 (2017).
\bibitem{reno-13}  M. Y. Pac (RENO Collab.), arXiv:1801.04049v1 [hep-ex].
\bibitem{dchoo-13} Y. Abe et al. (Double Chooz Collab.), Phys. Lett. B 735, 51 (2014).
\bibitem{icube-23} M. G. Aartsen et al. (IceCube Collab.), Phys. Rev. D 91, 072004 (2015).
\bibitem{nova-atm} P. Adamson et al. (NOvA Collab.), Phys. Rev. Lett. 118, 151802 (2017).
\bibitem{t2k-cp} K. Abe et al. (T2K Collab.), Phys. Rev. Lett. 118, 151801 (2017).
\bibitem{nova-morder} P. Adamson et al. (NOvA Collab.), Phys. Rev. Lett. 118, 231801 (2017).

\bibitem{flavor} Z.-Z. Xing, Int. J. Mod. Phys. A 19, 1 (2004).
\bibitem{flavor1} S. Luo and Z.-Z. Xing, Int. J. Mod. Phys. A 27, 1230031 (2012). 

\bibitem{mt} P. F. Harrison and W G Scott, Phys. Lett. B, 547 219 (2002).
\bibitem{mt1} C. S. Lam, Phys. Lett. B 507, 214 (2001).
\bibitem{mt2} H. Nishiura, K. Matsuda and T. Fukuyama, Int. J. Mod. Phys. A 23, 4557 (2008).
\bibitem{dsym} S. F. King and C. Luhn, Rep. Prog. Phys. 76, (2013).

\bibitem{masaki} I. Aizawa and M Yasue, Phys. Lett. B 607, 267 (2005).
\bibitem{masaki1} I. Aizawa, T Kitabayashi and M Yasue, Phys. Rev. D 72, 055014 (2005).
\bibitem{masaki2} T. Kitabayashi and M. Yasue, Phys. Lett. B 621, 133 (2005).
\bibitem{xing} Z.-Z. Xing and Y. L. Zhou, Phys. Lett. B 693, 584 (2010).
\bibitem{jora} R. Jora, J. Schechter and M. N. Shahid, Mod. Phys. Lett. A 28, 1350184 (2013).
 
\bibitem{bm} V. D. Barger, S. Pakvasa, T. J. Weiler and K. Whisnant, Phys. Lett. B 437, 107 (1998).
\bibitem{tbm} P. F. Harrison, D. H. Perkins and W. G. Scott, Phys. Lett. B 530, 167 (2002).

\bibitem{maxml} K. S. Babu, E. Ma and J. W. F. Valle, Phys. Lett. B 552, 207 (2003).
\bibitem{maxml1} W. Grimus and L. Lavoura, Phys. Lett. B 579, 113 (2004).
\bibitem{maxml2} X.-G. He, arXiv:1504.01560v3 [hep-ph].
\bibitem{maxml3} T. Fukuyama, arXiv:1701.04985v1 [hep-ph].
\bibitem{maxml4} Z.-Z. Xing and Z.-H. Zhao, Rep. Prog. Phys. 79, (2016).
\bibitem{maxml5} E. Ma, A. Natale and O. Popov, Phys. Lett. B 746, 114 (2015). 
            

              
\end{thebibliography}
\end{document}